# Revealing quantum path details in high-field physics

G. Kolliopoulos[1,2], B. Bergues[3], H. Schröder[3], P. A. Carpeggiani[1,2], L. Veisz[3], G. D. Tsakiris[3], D. Charalambidis[1,2] and P. Tzallas[1*]

[1]*Foundation for Research and Technology - Hellas, Institute of Electronic Structure & Laser, PO Box 1527, GR71110 Heraklion (Crete), Greece*

[2]*Department of Physics, University of Crete, PO Box 2208, GR71003 Heraklion (Crete), Greece*

[3]*Max-Planck-Institut für Quantenoptik, D-85748 Garching, Germany*

[*]Corresponding author e-mail address: ptzallas@iesl.forth.gr

**Abstract**

The tunnelling of an electron through a suppressed atomic potential followed by its motion in the continuum, is the fundamental mechanism underlying strong-field laser-atom/molecule interactions. Due to its quantum nature, the interaction is governed by the phase of the released electron wave-packets. Thus, detailed mapping of the electron wave-packet interferences provides essential insight into the physics underlying the interaction. A process which is providing access to the intricacies of the interaction is the generation of high order harmonics of the laser frequency. The phase-amplitude distribution of the emitted extreme-ultraviolet (EUV), carries the complete information about the harmonic generation process and vice-versa. Thus, the visualization of the EUV-spatial-amplitude-distribution, as it results from interfering electron wave-packet contributions, is of crucial importance. Restrictions to this accomplishment are due to the spatially integrating measurement approaches applied so far, that average out the phase effects in the generation process. In this work, we demonstrate a method which overcomes this obstacle. An EUV-spatial-amplitude-distribution-image is induced from the imprint on the measured spatial distribution of ions, produced through EUV-photon ionization of atoms. Interference extremes in the image carry phase information about the interfering electron wave-packets. The present approach provides detailed insight on the strong-field laser-atom interaction mechanism, while establishing the era of phase selective interaction studies. Furthermore, it paves the way for substantial enhancement of the spectral and temporal precision of measurements by *in-situ* controlling the phase distribution of the emitted radiation and/or spatially selecting the EUV-radiation-atom interaction products.

The demand for in depth understanding of processes occurring on the atomic level has stimulated the development of high precision metrology tools, which can be used for monitoring the fundamental properties of the matter on the quantum mechanical level. Coherent light sources, in the visible-UV and EUV spectral region, are efficiently serving this goal via light-matter interaction processes. Strong-field light-matter interactions induced by intense laser sources led to the observation of a broad range of phenomena such us the generation of coherent EUV radiation and the attosecond pulse formation [1-5] (and references therein). Recently, high-order-harmonics have been used for imaging atomic [6] and molecular orbitals [7-9] while the use of attosecond pulses pushed the temporal resolution of ultrafast dynamical studies towards the scale of the atomic unit of time. New pulse characterization techniques have been developed for measuring the duration of these pulses [1, 3], which have been exploited in the observation of a number of new processes in all states of matter [3, 4, 10-12]. On the other hand, frequency combs have been recently developed in the EUV spectral region [13-15], pushing the frequency resolution in the MHz range, improving thus the precision of measurement by an order of magnitude [14, 15]. Parallel to the fascinating results achieved so far by the use of EUV sources, studies of the strong-field laser-atom/molecule interaction mechanism are becoming more detailed [16, 17] since the mechanism is associated with the fundamental physical properties of the medium used for the EUV generation. Also it is directly linked to the coherence properties of the generated EUV radiation, which in turn, sets the upper limits of precision of the measurements performed with it.

The phase distribution of the EUV radiation emitted by the non-linear interaction of *fs* laser radiation with gases is associated with the high-order-harmonic generation process itself. The process at the single atom level, is governed by the electron quantum path interference [18 and references therein-22]. The properties of the emitted harmonics strongly depend on the intensity of the driving field ($I_L$). This dependence is associated with the phase accumulated by the electron wave-packets during their motion in the continuum. In the plateau spectral region, two intensity dependant quantum interfering electron trajectories, the *Long* (L) and the *Short* (S), with different flight times contribute mainly to the off-axis and on-axis harmonic emission [23] with phases $\varphi_q^L(I_L)$ and $\phi_q^S(I_L)$, respectively. In the deep cut-off spectral region the two trajectories degenerate to one with a single phase.

Elaborating on this matter, the image of the focused EUV beam (harmonics from 11$^{th}$ up to 15$^{th}$), was calculated for two different values of $I_L$. Figure 1a shows the profiles of the plateau harmonics on the surface of the focusing mirror. The outer part of the beam contains mainly radiation resulted by the "*Long*" electron trajectories while the inner part by the "*Short*" [23]. Fig 1b and 1c show the images of the focused harmonics generated at two $I_L$ values, for which harmonic phase distributions lead to on axis (*z*) destructive (Fig. 1b) and constructive (Fig. 1c) interference at the focus position with a visible double and single peak structure, respectively. The single and the double peak structures are resulting when the phase difference between the trajectories is $\Delta\varphi_q^{S,L} \approx 2n\pi$ and $\Delta\varphi_q^{S,L} \approx (2n+1)\pi$, respectively (*n* = 0,1,2…). The experimental observation of this interference pattern is the key point of the present study since it provides the spatial EUV-amplitude-distribution and spatial positions at which the "long"-"short" trajectory phase differences become $2n\pi$ or $(2n+1)\pi$ in the interaction area.

Here, we demonstrate an approach, with which we can image and control the above mentioned spatial intensity distribution of the EUV radiation generated by the non-linear interaction of Ti:Sapphire *fs* laser pulse with the Xenon gas (Fig. 2a). Argon and Helium gases were introduced in the interaction volume and have been single- and two-photon ionized, respectively by the 11$^{th}$ to 15$^{th}$ harmonics. Figures 2b, c show the ion distributions recorded at the EUV focus, at two different values of $I_L$ using Argon as a target gas. The characteristic interference patterns, exhibiting a double (Fig. 2d) and single peak (Fig. 2e) structure around the focus, are in agreement with the characteristic features of Fig's 1b and 1c, respectively. The structures are better discernable after subtraction of the smooth part of the signal (Fig. 2f, g).

As the electron quantum path interference is strongly dependent on $I_L$, we have performed a systematic recording of the ionization focus images at different driving intensities. This dependence is shown in the contour plot of Fig. 3a. In this measurement, an intensity dependant interference pattern along the *z* axis was observed. This dependence is clearly shown as a modulation from a single to a double peak structure in the normalized contour plot of Fig. 3b, which clearly confirms the statistical relevance of the structures shown in Figs. 2f and 2g. The "saw-type" structure is the result of the phase distribution of the harmonics due to the spatial dependence of the electron quantum path interference in the harmonic generation process. This structure is strongly pronounced at high $I_L$ values, where most of the harmonics lay in the plateau region. It is also observable at lower $I_L$ values where a transition of the harmonics from the plateau to the cut-off region takes place. A clear and intensity independent, single peak structure appears only at very low $I_L$ values, at which all the harmonics are clearly in the deep cut-off region. The measured structures are found to be in fair agreement with those retrieved from the single-atom three-step quantum mechanical model depicted in (Fig. 3c) (Supplementary Material).

The measured "saw-type" structure provides a direct access to the fundamental mechanism underlying the harmonic generation process (Fig. 4a). Although the harmonic generation mechanism has been extensively studied in the past [3, 16-19, 23, 24], a direct measurement of the difference between harmonic emission times ($\Delta t_e = t_e^{L_q} - t_e^{S_q}$) and electron quantum paths ($\Delta L_e = L_e^{L_q} - L_e^{S_q}$) resulted by the "*Long*" and "*Short*" trajectories, has never been implemented. Here, the values of $\Delta t_e$ have been deduced from the "saw-type" structure by using the following considerations: a) $\Delta \varphi_q^{S,L} = 0$ at lower values of $I_L$, where an $I_L$ independent single peak structure is observed. This is in agreement with the harmonic generation theory, where in the cut-off region the two trajectories degenerate to one with a single phase. b) $\Delta \varphi_q^{S,L}$ is increasing monotonically with $I_L$ [24]. c) The $\Delta \varphi_q^{S,L}$ is increasing by $\pi$ when the structure changes from a single to a double peak. Following these considerations, the values of $\Delta \varphi_q^{S,L}$ with differences $n\pi$ ($n=0,1,2…$) and $\Delta t_e = \Delta t_e^q = nT_q/2$ as a function of $I_L$, have been obtained and are shown in Figs 4b and 4c, respectively. $T_q/2$ is the half period of the $q$th harmonic. At $I_L$ regions where more than one harmonics lies in the plateau (Supplementary Material), $\Delta t_e$ reflects average value of the emission time differences weighted by the harmonic amplitude ($E_q$), i.e. $\Delta t_e = \langle \Delta t_e^q \rangle = \sum_q E_q \Delta t_e^q / \sum_q E_q$. In terms of electron path length, the difference between the trajectories can be obtained by the relation $\Delta L_e = \sqrt{2(E_e + IP)/m_e} \Delta t_e$ (where $E_e + IP = \hbar \omega_q$ is the final kinetic energy of the recolliding electron, $E_e$ is the return energy of the electron, and $IP$ is the binding energy of the atom) [16, 18]. As for a single recollision $0 \leq E_e \leq 3\, U_p$ [18], the above relation can be approximated by $\Delta L_e \approx \sqrt{2(1.5 \times U_p + IP)/m_e}\, \Delta t_e$ ($U_p$ is the ponderomotive energy of the electron). An interesting observation in the measured data is the correlation of the emission time differences with the electron quantum path interferences. Extracting the ratio $f(I_L) = \Delta L_e/\lambda_e$ of $\Delta L_e$ to the electron De Broglie wavelength ($\lambda_e = h/\sqrt{2m_e(1.5 \times U_p + IP)}$) (Fig. 4d), it has been found that for emission time differences $\Delta t_e \approx n\mathrm{T}_q/2$, the electron quantum path difference is changing by $\Delta L_e \approx n\lambda_e$. This is clearly shown on the intensity dependence of the differences between the consecutive values of $f(I_L)$ ($\Delta f(I_L) = f(I_L^{(i)}) - f(I_L^{(i+1)})$, $i=1..12$) which results to $\Delta f(I_L) \approx 0$ and $\Delta f(I_L) \approx \lambda_e$ for the cut-off and the plateau, respectively (inset of Fig. 4d). The small drift with the intensity of $\Delta f(I_L)$ from mean value $\lambda_e$, can be attributed to deviation of $E_e$ from the value of $1.5 U_p$.

Extending our technique to non-linear probing schemes using a target gas with a higher ionization potential could further enhance the precision of the measurements. Initial experiments in which helium is used to image the EUV-focus via a two-photon non-resonant ionization process (see Supplementary Material) demonstrate that a big step towards this goal has already been achieved.

In the present work, a direct access to the harmonic generation process has been achieved by mapping the spatial EUV-yield-distribution onto spatial ion distribution, produced in the EUV focal area through linear and non-linear processes of atoms. Manipulating the intensity dependant electron quantum paths, a direct measurement of the "short"-"long" trajectory harmonic emission time and electron quantum path differences have been obtained, providing at the same time a quantitative correlation between the photoemission time difference and the electron quantum path interferences. The approach can shed light on several strong-field light-matter interaction processes resulting in a stimulated light emission, including harmonic generation from atoms/molecules, surface plasma and bulk crystals [7-9, 25-29]. Attosecond science [1-5, 10-12], high resolution spectroscopy studies in the EUV [13-15] spectral region and molecular tomography methods [7-9], are some of the research topics that can markedly benefit from the *in-situ* control of the emitted

EUV phase distribution and/or the spatial selection of the EUV-radiation-atom interaction products.

**Acknowledgments**

This work is supported in part by the European Commission programs ATTOFEL, CRISP, Laserlab Europe, the Greek funding program NSRF and performed in the framework of the IKYDA program for the promotion of scientific cooperation between Greece and Germany.

**Figure captions**

**Figure 1. Calculated EUV focus images. a)** Profiles of harmonics 11[th], 13[th] and 15[th] on the surface of a spherical mirror of 5 cm focal length (used in order to focus the EUV beam), at laser peak intensity of $I_L=I_{max}=10^{14}$ W/cm$^2$. The relative amplitudes and phases of the inner and outer beams where calculated using the single-atom three-step model. The panels in **b)** and **c)** show the calculated images of the focused EUV beam, at two different values of $I_L$. At these intensities, the phase distribution of the EUV beam lead to on axis (*z*) destructive (b) and constructive (c) interference at the focus position with a double (maxima at z = ± 30 μm) and single (maximum at z = -8 μm) peak structure, respectively. Below each panel the intensity line-outs along the propagation axis (*z*) is shown. Similar features appear in the image of the focal area for each individual harmonic. The interference pattern appearing in the orange shaded areas is the fingerprint of the spatial phase distributions of the beam along the *z* (propagation) axis. The phase distribution of the harmonics for each $I_L$, has been calculated by using the three-step quantum mechanical model (Supplementary Material). At these values of $I_L$, the harmonics 11[th], 13[th] and 15[th] are lying in the plateau spectral region having approximately equal relative amplitudes.

**Figure 2. Apparatus used for imaging of the EUV-phase-distribution. a)** Intense EUV radiation [9, 11] is generated in Xenon gas by focusing, with a spherical lens (L) of 3m focal length, an IR 33 *fs* long laser pulse onto a Xenon gas jet (P-GJ). The EUV radiation is reflected towards the interaction region by a Silicon plate (SP) which has been used for the elimination of the IR beam. "F", "A" and "D" are the filter, aperture and the calibrated EUV photodiode, respectively. The transmitted harmonics 11[th], 13[th] and 15[th] were focused by an unprotected gold spherical mirror of focal length *f* = 5 cm (SM) onto a second gas jet with the target gas Argon or Helium (T-GJ). The images of the EUV focus, projected orthogonally to the beam propagation direction (z), were monitored by means of a high spatial resolution ion-imaging-detector (I-ID) [30]. **b)** and **c)** EUV focus images recorded by monitoring the single-photon ionization signal of Argon, at intensity values $I_L$ ≈ 0.73 x $I_{max}$ and ≈ 0.67 *x* $I_{max}$, respectively. 600 shots were accumulated for each image. The relative amplitudes of the harmonics 11[th], 13[th] and 15[th] in the interaction region were approximately 0.9:1:0.7, for both images. In this measurement the EUV beam was spectrally filtered by means of a thin Sn filter.

**d)** and **e)** The characteristic interference pattern along the *z* axis, with a double (maxima at z = ± 32 µm) and single (maximum at z ≈ -12 µm) peak structure around the focus, respectively. To make the trend clearer, the interference pattern in **f)** and **g)** retrieved after subtraction of the smooth part of the signal (green line) in (d) and (e), respectively. The main reason for the limited modulation depth comes from the contribution of the out of plane signal in the projected focus image, as well as from the finite resolution of the I-ID. The red lines in figs **d-g** correspond to a smoothing of 30 points on the raw data. The error bars represent one standard deviation of the mean value of the 30 points.

**Figure 3. Visualization of the intensity dependant spatial EUV-yield-distribution. a)** Contour plot which shows the dependence of the interference pattern (induced by single-photon ionization of Argon) along the *z* axis on the $I_L$. For each value of $I_L$, 600 shots were accumulated. The focus position of the IR field was placed on the Xenon gas jet. In order to avoid any influence on the EUV beam that may introduced by the filter, a spectrally unfiltered EUV radiation has been used. The relative amplitudes of the harmonics $11^{th}$, $13^{th}$ and $15^{th}$ in the interaction region are shown in the Supplementary Material. **b)** Contour plot retrieved after normalization of the plot (a). The black line depicts the mean value of the ion distribution. The error bars represent the standard deviation of the mean value resulted by the accuracy on measuring the laser intensity. **c)** Normalized calculated contour plot which shows the dependence of the interference pattern along the z axis on the $I_L$. Although, the "saw-type" structure and the position of the maxima on the z axis are in agreement with the measured values, these calculations are meant to show the general features of such measurements and do not provide a quantitative comparison.

**Figure 4. Direct measurement of harmonic emission time and electron quantum path differences. a)** High-order-harmonic generation mechanism. The IR field suppresses the atomic potential and allows the valance electron to tunnel through. The electron moves almost freely in the driving field gaining kinetic energy, which is converted to photons upon its recombination. Due to the different electron trajectories the harmonic emission times are different. $t_e^S$ and $t_e^L$ correspond to the harmonic emission times resulted by the "*Short*" and "*Long*" trajectories, respectively. **b)**

Measured phase difference $(\Delta\varphi_q^{S,L})$ between the "Short" and "Long" trajectories as a function of $I_L$. **c)** Difference between the emission times $(\Delta t_e)$ of the harmonics resulted by the "Short" and "Long trajectories. The values of $\Delta t_e$ retrieved from the measured phase differences $(\Delta\varphi_q^{S,L})$. **d)** Correlation of the emission time differences with the electron quantum path interferences. Comparison of $\Delta L_e$ with De Broglie electron wave $\lambda_e$ via the relation $f(I_L)=\Delta L_e/\lambda_e$. The inset shows the difference between the consecutive values (where the $\Delta\varphi_q^{S,L}$ increment is $\pi$) of $f(I_L)$ ($\Delta f(I_L) = f(I_L^{(i)}) - f(I_L^{(i+1)})$, $i$=1…12). For emission time differences $\Delta t_e \approx nT_q/2$ the electron quantum path differences are changing by $\Delta L_e \approx n\lambda_e$ (the expression is shown in the upper panel of Fig. 4a). The vertical black lines depict the harmonic cut-off regions. The error bars represent one standard deviation of the mean value.

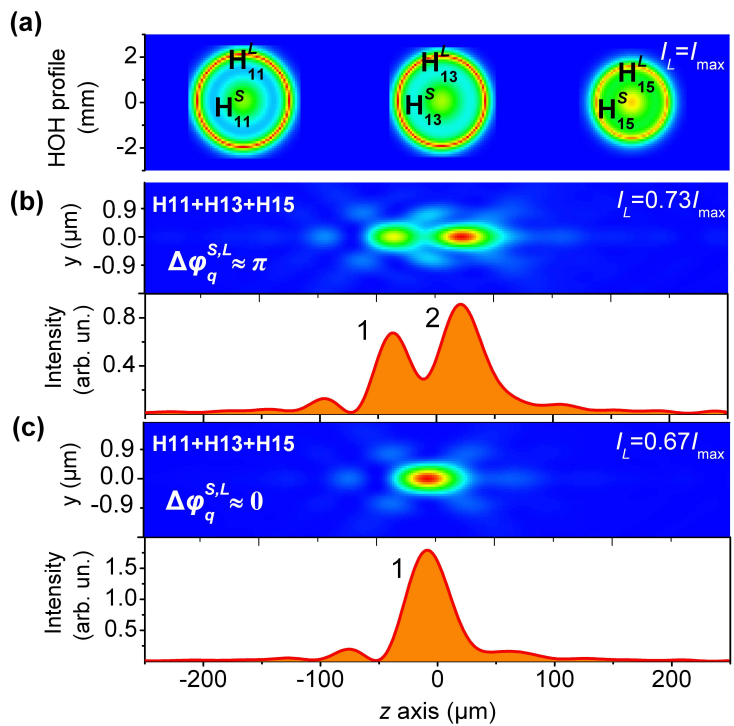

**Figure 1.** *G. Kolliopoulos et al.*

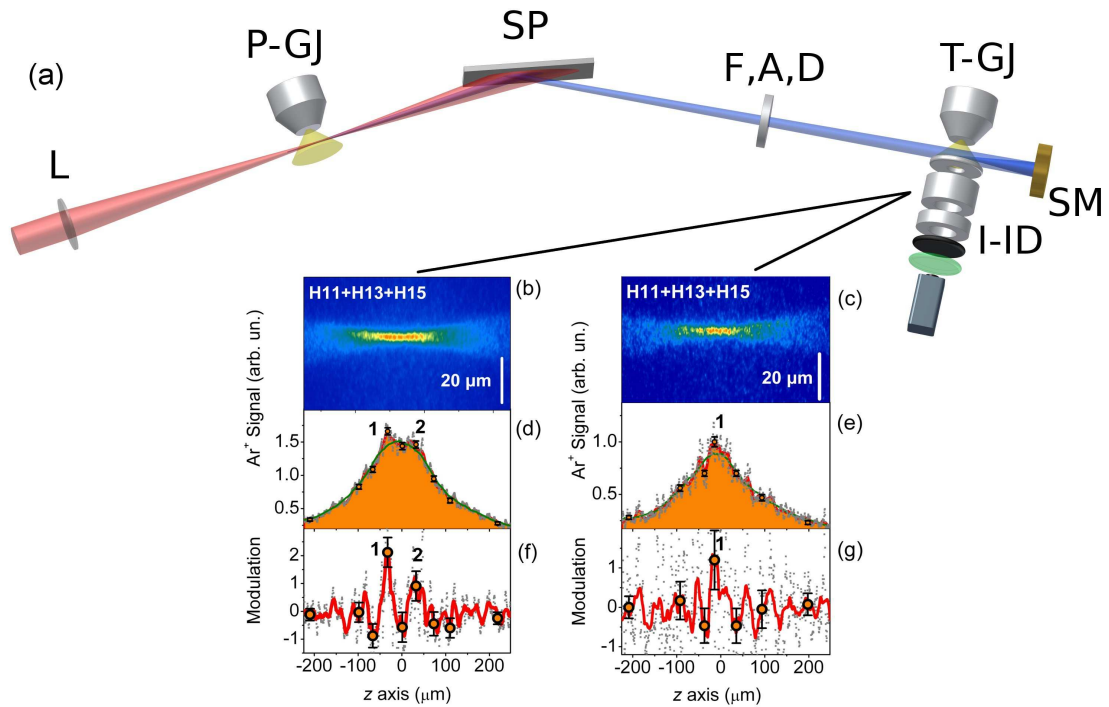

**Figure 2.** *G. Kolliopoulos et al.*

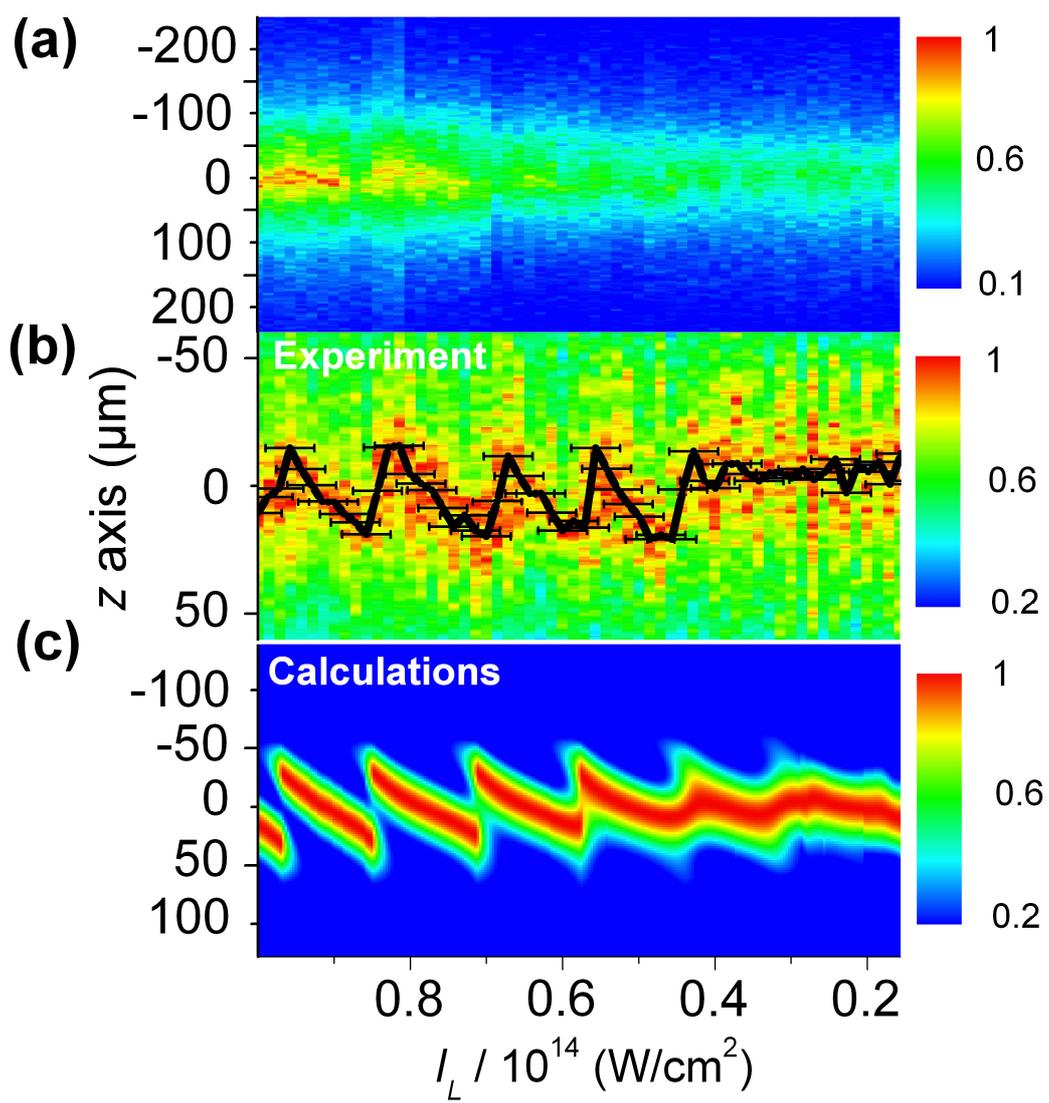

**Figure 3.** *G. Kolliopoulos et al.*

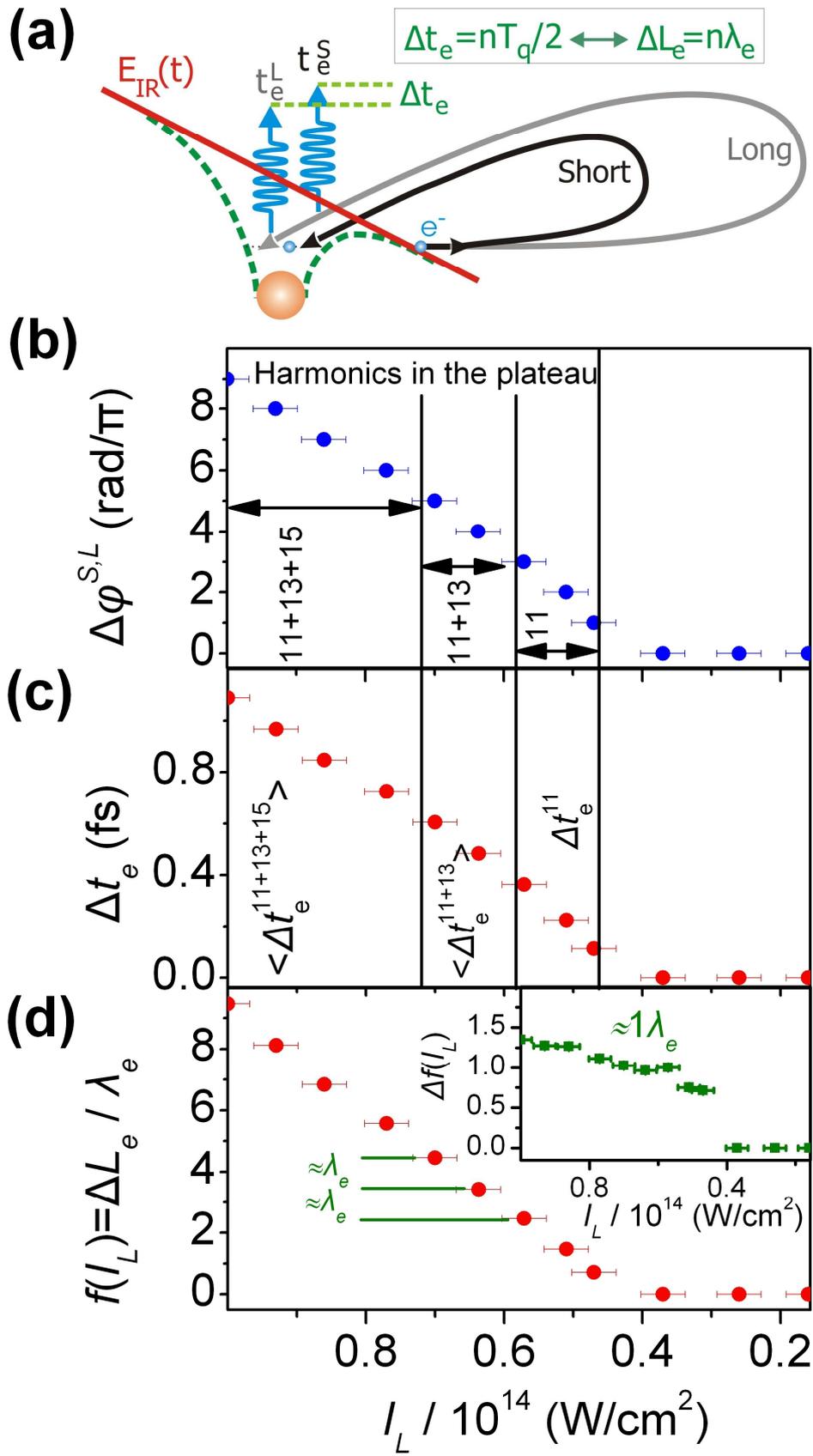

**Figure 4.** *G. Kolliopoulos et al.*

Supplementary Material

**Experimental procedure**

The experiment was performed utilizing a 10 Hz repetition rate Ti:sapphire laser system delivering pulses of up to 170 mJ energy, $\tau_L$=33 fs duration and wavelength at 800 nm (IR). The experimental setup, which partially is shown in Fig.2a of the manuscript, can be seen shown in Fig.SI-1.

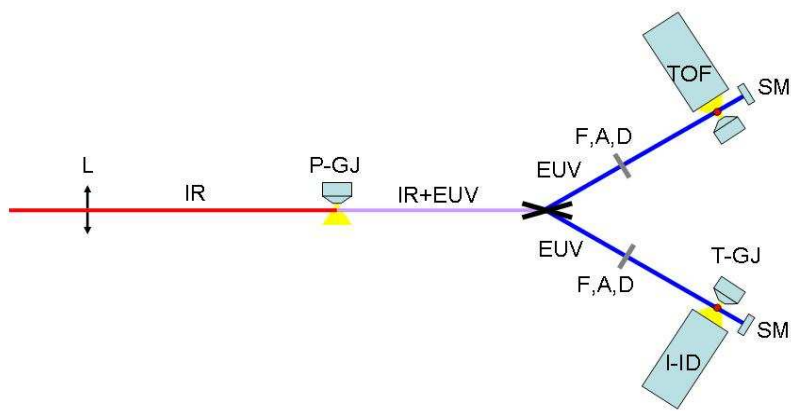

**Figure SM-1. Experimental set-up.** (L): Spherical lens for focusing the IR beam into the Xe gas jet (P-GJ). (SP): Silicon plate which reflects the EUV beam towards the interaction area. (F, A): Sn filter and Aperture used for the spectral and spatial filtering. (D): EUV-calibrated photodiode used for the measurement of the EUV energy. (SM): Unprotected spherical gold mirror used for focusing the EUV beam into a second gas jet filled with the target gas (T-GJ). (I-ID): Ion-Imaging-Detector used for monitoring the EUV images. (TOF): Time of flight electron spectrometer used for the measurement of the EUV spectrum.

An annular laser beam (formed using a super-Gaussian beam stop) with outer diameter of ≈2.5 cm and energy of ≈15 mJ/pulse was focused with an $f$ =3 m lens into a pulsed gas jet (P-GJ), filled with Xe where the harmonic radiation was generated. For the measurements presented in the figures of the manuscript the focus position of the IR beam was fixed at the Xe gas jet. After the jet a Si plate was placed at Brewster's angle of the fundamental (75°), reflecting the harmonics [1] towards the detection area, while substantially reducing the IR field. The EUV radiation after reflection from the Si plate passes through a 5 mm diameter aperture (A) which blocks the residual outer part of the IR beam.

Subsequently the EUV beam was focused into the target gas jet (T-GJ) by a spherical gold mirror (SM) of 5 cm focal length. Great care has been taken on fixing the angle of incidence of the EUV beam on the gold mirror at zero degrees. The EUV images were monitored by means of a high spatial resolution (1 μm) ion-imaging-detector (I-ID) [2, 3], which records the spatial distribution of the ionization products resulted by the interaction of the EUV light with the target gas.

The spectrum of the EUV radiation in the interaction area, was determined by measuring the energy-resolved, single-photon ionization, photoelectron spectra of Ar gas. The electron spectra were recorded using a μ-metal shielded time-off-flight (TOF) ion/electron spectrometer, attached to a second EUV beam-line brunch (note that the TOF branch is not shown in Fig. 2a of the manuscript). In order to have the same experimental conditions in both, the TOF and the I-ID set-ups, the branch of TOF, was constructed in an identical way as the one of the I-ID. This is done by using, a second Si plate (mounted on a translation stage) placed at Brewster's angle of the fundamental (75°). The length of the second branch, the metal filter, the aperture and the spherical gold mirror were the same as those used in the I-ID branch. For the measurement of the photoelectron spectrum in the interaction region the focused EUV beam was tilted by ≈1.5 deg with respect to the incoming beam. This is done in order to avoid measuring the photoelectrons resulted from the incoming EUV beam. For the EUV images shown in Fig. 2 of the manuscript and Fig.SI-3, an 150 nm thick Sn filter, which transmits only the harmonics $11^{th}$, $13^{th}$ and $15^{th}$, has been placed after the aperture. The amplitudes of the harmonics $11^{th}$, $13^{th}$ and $15^{th}$ in the interaction area were found to be in arbitrary units 0.9, 1 and 0.7, respectively. For the images shown in Fig. 3 of the manuscript, the Sn filter was taken out of the EUV beam line, in order to avoid any influence on the properties of the EUV beam which may introduced by the filter. In this case, while all the generated harmonics are entering into the interaction area only the harmonics above the $11^{th}$ can ionize the Ar gas by single-photon absorption. The peak heights of the harmonics $11^{th}$, $13^{th}$ and $15^{th}$ after the reflection from the spherical gold mirror are shown in arbitrary units in Fig. SI-2c. The dependence of the harmonic signal on $I_L$, has been used for the measurement of the $E_q$ and the determination of the cut-off regions.

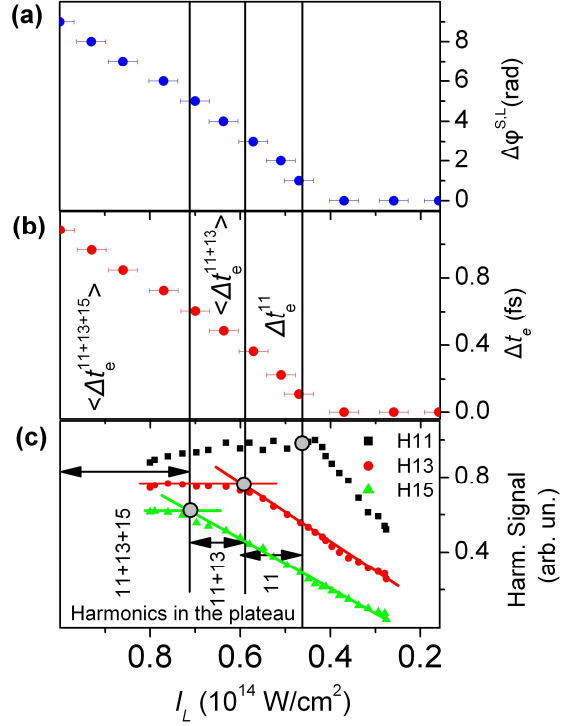

**Figure SM-2. Direct measurement of harmonic emission time differences. a)** Measured phase difference $(\Delta\varphi_q^{S,L})$ between the "Short" and "Long" trajectories as a function of $I_L$ **b)** Difference between the emission times $(\Delta t_e)$ of the harmonics resulted by the "Short" and "Long trajectories. The values of $\Delta t_e$ retrieved from the measured phase differences $(\Delta\varphi_q^{S,L})$. **c)** The dependence of the harmonic signal on $I_L$. The harmonic signal has been recorded up to $I_L$ values where all the harmonics are in the plateau region. The vertical black lines depict the harmonic cut-off regions.

The 17th harmonic has been detected (at high $I_L$) with an order of magnitude smaller amplitude than the 11th harmonic. All higher harmonics higher than 17th have even smaller amplitudes and thus are negligible. The relative amplitudes of the harmonics 11th, 13th and 15th in the interaction region were found to be approximately the same with those of the spectrally filtered EUV radiation. The energy of the EUV radiation has been measured by means of an XUV calibrated photodiode which has been placed after the aperture (A). At the maximum laser intensities ($I_{Lmax} \approx 10^{14}$ W/cm$^2$) used, the outer 4.6 mm EUV beam diameter on the focusing gold mirror and the 2±1 μm EUV focal spot diameter have been measured using the I-ID. To the ±1 μm error in the measured focal spot diameter contribute the spatial resolution of the I-ID and the shot-to-shot point stability of the EUV at the focus which found to be 1 µm. Utilizing phase matching conditions favourable for "long trajectory harmonic emission" (IR focus before the Xe gas jet) it has been found that the divergence of the EUV beam is 2 times larger than that of "short trajectory harmonic emission" (IR focus before the Xe gas jet). For the measurement of the EUV image resulted by the 2-photon-ionization of He (Fig. SI-3) the EUV energy has been maximized and was found to be ≈120 nJ in the interaction area. Taking into account the energy, the measured focal spot diameter and assuming an average EUV pulse duration of 15 fs, the intensity of the EUV pulse in the He gas jet is estimated to be ≈$5 \times 10^{14}$ W/cm$^2$.

**Two-photon ionization images**

Using Helium as a target gas, the EUV focus image was monitored by recording the ions produced by a two-photon non-resonant ionization process. Figs' SI-3a and b show the ion distributions recorded at the EUV focus, at two different values of $I_L$. An obvious difference between the single- and two-photon ionization images, which is associated with the different order of non-linearity of the involved processes, is the overall width of the contrast of the ion distribution pattern. The width of the profile along the z axis is found to be a factor of ≈1.4 narrower in Helium than in Argon. Although the noise doesn't ensure the existence a clear interference pattern, the broader peak of the image shown in Fig. SI-3a (Fig. SI-3c) compared to the image shown in Fig. SI-3b (Fig. SI-3d), together with the indication of the existence of a double and a single peak structures, show the applicability of the scheme on performing experiments with enhanced precision. Furthermore, it can be seen as an ideal tool towards a single-shot attosecond pulse metrology. A single-shot 2$^{nd}$ order autocorrelation can be performed by producing such images using two EUV replicas with different wave-vector projections along the axis at which the measurement is taking place.

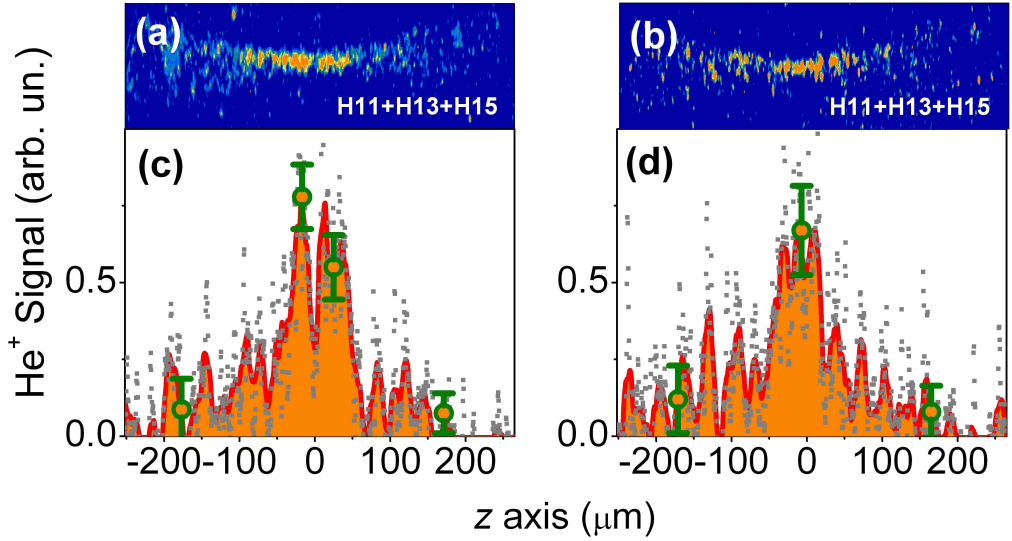

**Figure SM-3. Non-linear visualization of the EUV-amplitude-distribution. a)** and **b)** EUV focus images recorded by monitoring the two-photon ionization signal of Helium, at intensity values of $I_L \approx I_{max}$ and $\approx 0.9 \times I_{max}$, respectively. The EUV radiation was spectrally filtered by using an 150 nm thick Sn metal filter. For each image 15000 shots were accumulated. **c)** and **d)** Intensity line-outs along the propagation axis (z).

**Theoretical procedure**

The coherence of the EUV radiation emitted from gases, is associated with the high-order-harmonic generation process, which in the single atom frame, is governed by the electron quantum path interference [4-7] and in the macroscopic scale, by the phase-matching conditions [8-12]. The spatiotemporal coherence properties of the emitted harmonics strongly depend on the harmonic order ($q$), the intensity of the driving field ($I_L$), the focusing conditions, and the dispersion properties of the medium. The dependence on $I_L$, is associated with the phase accumulated by the electron wave-packets during their motion in the continuum. In the single atom frame, two quantum interfering electron trajectories, the *Long* (L) and the *Short* (S), with different flight times $\tau_q^L(I_L)$ and $\tau_q^S(I_L)$, contribute to the emission of each harmonic $q$ in the plateau spectral region. In the deep cut-off spectral region the two trajectories degenerate to one with a single phase for the cut-off harmonic. The electron wave-packet phases, and with those the phases of the emitted harmonics ($\varphi_q^{L,S}$), are approximately proportional to the product $\tau_q^{L,S}(I_L)I_L$. Also, due to the phase matching conditions the harmonics resulted by the long trajectories have larger divergence

(strong off-axis emission) as compared to the harmonics resulted by short trajectories (strong on-axis emission). The overall divergence of the harmonic beam is reduced with the harmonic order [13, 14].

The calculated EUV focus images presented in the manuscript are obtained by the Debye integral [15], after applying the Huygens–Fresnel principle on a spherical mirror with 10 cm radius of curvature. The spectral phase distribution and the relative amplitudes between the "short" and "long" trajectory harmonics, have been calculated by using the single-atom three-step quantum mechanical model [4]. The relative amplitudes between different harmonics have been taken from the graph SI-2c. In the calculations, for laser intensities in the plateau, the beam diameters on surface of the mirror for the harmonics emitted by the "short" and "long" trajectories were determined by means of I-ID and knife-edge technique (at different positions of the IR focus relative to the gas jet). When the harmonics enter in the cut-off region the beam diameter reduces according to the relation $\tilde{d}_s(p) = d_s / \sqrt{p}$ and $\tilde{d}_L(p) = d_L / \sqrt{p}$ ( $p = q - q_{cut-off}$ ) (where $q_{cut\text{-}off}$ is the cut-off value for each $I_L$). This approximation is based on previous experimental and theoretical studies [13, 14, 16, 17]. The profile of each harmonic is obtained by superposing the harmonic fields resulted by the emission of the "short" and "long" trajectories.